\documentclass{mem}
\usepackage{natbib}\usepackage{txfonts}\usepackage{balance}
\usepackage{graphicx}
\usepackage[a4paper]{hyperref}
\idline{75}{282}
\begin{document}
\def\mincir{\raise -2.truept\hbox{\rlap{\hbox{$\sim$}}\raise5.truept \hbox{$<$}\ }}
\def\magcir{\raise-2.truept\hbox{\rlap{\hbox{$\sim$}}\raise5.truept \hbox{$>$}\ }}

\title{VHE astrophysics: recent developments}

   \subtitle{}

\author{
Massimo Persic\inst{1}
\and 
Alessandro De~Angelis\inst{2}
          }

\offprints{M.~Persic}

\institute{
INAF and INFN, Trieste, Italy ~~~(\email{persic@oats.inaf.it})
\and
Universit\`a di Udine and INFN, Udine, Italy
}

\authorrunning{Persic \& De~Angelis}

\abstract{
We review the current status, and some open issues, of VHE astrophysics. 
\keywords{Gamma rays: observations, supernova remnants, pulsars, pulsar wind nebulae, 
galaxies, active galactic nuclei; cosmic rays; dark matter }
}

\maketitle{}

\section{Introduction}

The ground-breaking work of the imaging atmospheric Cherenkov telescope (IACT) Whipple
led to the earliest detections of sources in the VHE (i.e., $\sim$0.1-100~TeV) band,
the Crab pulsar and nebula (Weekes et al. 1989) and the blazar Mrk~421 
(Punch et al. 1992). These pioneering attempts initiated VHE astrophysics. Following 
on, the first generation of major IACTs -- that included CAT (1996-2003), CANGAROO 
(1992-2001) and HEGRA (1993-2002), in addition to Whipple (1969-2003) itself -- 
broadened the discovery potential of the new field by detecting several more blazars 
and Galactic sources.
Thanks to reduced low-energy thresholds, improved sensitivities, wider-field cameras, 
and lighter mechanical structures, the current second-generation IACTs -- i.e., H.E.S.S. 
(2003-), MAGIC (2004-), CANGAROO III (2004-), and VERITAS (2006-) -- have taken VHE 
astrophysics into maturity. 

In this paper we highlight some recent progress in VHE astrophysics obtained with the 
current generation of instruments.

\section{Galactic sources}

In a seminal VHE survey of the Galactic plane, the southern-located H.E.S.S. telescope discovered 
14 previously unknown sources (Aharonian et al. 2006a). Further Galactic sources, accessible from 
the northern hemisphere, were subsequently observed with the MAGIC telescope (e.g., Albert et al. 
2007a). Proposed counterparts of such Galactic VHE sources include supernova remnants (SNRs), pulsar 
wind nebulae (PWNe), and accreting binaries. Whatever their detailed nature, it is expected that 
Galactic VHE sources are related to evolutionary endproducts of massive, bright, short-lived, 
stellar progenitors. Hence, these Galactic VHE sources are immediate tracers of the current star 
formation. 

\subsection{Supernova remnants}

Galactic cosmic rays have long been suspected to be produced at supernova (SN) shock fronts via 
diffusive acceleration. If the observed VHE $\gamma$-rays were found to be generated through the 
hadronic channel, via $\pi^0$ decay following pp interaction with the dense molecular clouds embedding 
the short-lived SN progenitor, then the acceleration by SNe of nuclei to energies of the order of 
the knee in the CR spectrum would be virtually proven (e.g., Torres et al. 2003). However, it is 
difficult to disentangle the hadronic VHE component from the leptonic one, produced by inverse-Compton 
(IC) scattering of interstellar radiation field photons (in the inner Galaxy) or CMB photons (in the 
outer Galaxy) off ultrarelativistic electrons (e.g., Porter et al. 2006), by measuring $\gamma$-rays 
over only a decade or so in energy. The VHE data of RX~J1713.7-3946 can be explained in terms of 
either channel, leptonic/hadronic if the relevant magnetic field is low/high ($B$$\sim$10/100$\,\mu$G: 
Aharonian et al. 2006b and Berezhko \& V\"olk 2006). Data in the $\sim$0.1--100~GeV band, such as those 
to be provided by AGILE and GLAST, are clearly needed to discriminate between the two channels. (For 
Cas~A the high magnetic field, $B$$\sim$1~mG, suggests a mostly hadronic [Berezhko et al. 2003] VHE 
emission [O\~na-Wilhelmi et al. 2007].) 

Whatever the details, the detection of photons with energy $\magcir$100 TeV from RX~J1713.7-3946 is a 
proof of the acceleration of primary particles in SN shocks to energies well above 10$^{14}$~eV. The
differential VHE spectral index is $\sim$2.1 all across this SNR, suggesting that the emitting
particles are ubiquitously strong-shock accelerated, up to energies $\sim$200/100 TeV for primary CR
protons/electrons if the hadronic/leptonic channel is at work (Aharonian et al. 2007). This is getting 
close to the knee of the CR spectrum, $\sim$10$^{15.5}$~eV, that signals the high-energy end of the 
Galactic CR distribution (e.g., Blasi 2005).

Circumstantial evidence supports a hadronic origin of the VHE emission. In several expanding SNRs 
the X-ray brightness profile behind the forward shock is best explained as synchrotron emission from 
energetic electrons in high magnetic fields, $B$$\sim${\cal O}(10$^2)\,\mu$G, i.e. $\sim$100 times 
larger than typical interstellar medium (ISM) values. Such a large amplified magnetic field disfavors 
the IC interpretation of the VHE data. Furthermore, in the remnant H.E.S.S.$\,$J1834-087 the maximum of 
the extended VHE emission correlates with a maximum in the density of a nearby molecular cloud 
(Albert et al. 2006a) -- which suggests hadronic illumination of the target molecular cloud.

\subsection{Pulsars and associated nebulae}

Discrimination between different processes of pulsar magnetospheric emission (e.g., polar-cap vs
outer-gap scenario) is one clear goal of VHE astrophysics. Polar-cap (e.g., Daugherty \& Harding 
1982, 1996) and outer-gap (e.g., Cheng et al. 1986; Romani 1996) models essentially
differ by the location of the gap in the pulsar magnetosphere. In the former case
this is close to the neutron star (NS) surface, whereas in the latter it is further away from it.
Thus, the influence of the magnetic field ($B$$\sim$10$^{11-13}$~G) is crucially different in
these models. In polar-cap models, it produces absorption (due to $\gamma$$+$$B$ $\rightarrow$$e^\pm$)
leading to a super-exponential cutoff of the emission (mostly curvature radiation). In outer-gap
models, only a (milder) exponential cutoff is present, and the highest photon energies depend on
the electron energy: in these models a VHE IC component may thus arise from the upscattering, by
such energetic electrons, of their emitted synchrotron photons or of the X-ray photons released
by NS heating. Both models can deal with current observational constraints.

The current situation is exemplified be the cases of the Vela and PSR~B1951+32 pulsars 
(Aharonian et al. 2006c; Albert et al. 2007b). In the latter case upper limits to 
the pulsed emission imply a cutoff energy $E_c$$<$32~GeV.
In both, Vela and PSR~B1951+32, IC emission at TeV energies as predicted by outer-gap models is
severely constrained, although not all outer-gap models are ruled out. Deeper sensitivities can test
the models further, and certainly no test of these models in the range 10--100 GeV has yet been
achieved.

Millisecond pulsars, that have lower magnetic fields ($B$$\sim$10$^{8-10}$~G), in polar-cap models 
could have $E_c$$\magcir$100 GeV (i.e., $E_c$$\propto$$B^{-1}$), hence their (pulsed) VHE emission 
could be substantial. Detecting it would be a test of the polar-cap theory (Harding et al. 2005) -- 
however, no positive detection has been reported so far.

The Crab nebula, a steady emitter that is used as a calibration candle, has been observed 
extensively from the radio up to $\sim$70~TeV. No pulsed (magnetospheric) VHE emission was found 
in MAGIC data, that implies 
$E_c$$<$30~GeV (Albert et al. 2007c). The steady nebular spectrum, measured in 
the $\sim$0.03-30~GeV range by EGRET and in the $\sim$0.06-70~TeV by several IACTs, shows a bump 
that starts to dominate at $\sim$1~GeV and peaks at $\sim$50~GeV: this component results 
from IC scattering, by the synchrotron-emitting electrons, of softer photons in the shocked wind 
region -- i.e., synchrotron, FIR/mm or CMB photons. In spite of its detected IC $\gamma$-ray emission, 
however, the Crab nebula is not an effective IC emitter as a consequence of its high nebular magnetic 
field ($B$$\sim$0.4~mG). 

PWNe -- i.e., pulsars displaying a prominent nebular emission -- currently constitute the most populated class 
of identified Galactic VHE sources (7 identifications -- e.g., Gallant 2006). The VHE emission of PWNe 
is likely of leptonic origin. Let us consider the case of H.E.S.S.~J1825-137, for which spectra have been 
measured in spatially separated regions (Aharonian et al. 2006d). In these regions, the VHE spectra steepen 
with increasing distance from the pulsar, and the VHE morphology is similar to the X-ray morphology: 
furthermore the low derived magnetic field (few $\mu$G) implies that synchrotron X-ray emission is due to 
electrons of energy higher than the $\gamma$-rays. This suggests that the observations can be interpreted in 
terms of very energetic electrons that efficiently lose energy via synchrotron losses, aging progressively 
more rapidly as they are farther away from the acceleration site, and produce VHE $\gamma$-rays via IC 
scattering.

\subsection{TeV binaries}

In both SNRs and PWNe particle acceleration proceeds on the parsec distance scales in the 
shocks formed in interactions of either SN ejecta or pulsar winds with the ISM. A 
different population of much more compact particle accelerators, which has been revealed 
by current IACTs, is formed by the TeV binaries (TVBs). These systems contain a compact 
object --either a NS or a black hole (BH)-- that accretes, or interacts with, matter 
outflowing from a companion star: hence they are VHE-loud X-ray binaries (XRBs). Four TVBs 
have been detected so far: PSR~B1259$-$63 (Aharonian et al. 2005a), LS~5039 (Aharonian et 
al. 2005b, 2006e), LS~I~+61~303 (Albert et al. 2006b), and Cyg~X-1 (Albert et al. 2007d). 

PSR~B1259-63 is powered by the rotation energy of its young 48~ms pulsar, and strong VHE 
emission is observed from this system in pre- and post-periastron phases when the relativistic 
pulsar wind collides with the dense equatorial wind blowing from the companion Be star.  

LS~I~+61~303 and LS~5039 may share a similar structure. The former is composed of a compact 
object and a Be star in a highly eccentric orbit. Its VHE emission, whose variability 
constrains the emitting region of LS~I~+61~303 to be larger than the binary system's size, 
appears correlated with the radio emission and does not peak at periastron, where $\dot M$ 
is expected to be largest. This picture favors an IC origin of VHE emission, as probably the 
most efficient at the relatively large scales of the system at peak emission.

The emission from Cyg~X-1 is point-like and excludes the nearby radio nebula powered by the 
relativistic jet. Cyg~X-1 is the first stellar-mass BH, and hence the first established 
accreting binary, detected at VHE frequencies. 

\subsection{Galactic center}

The possibility of indirect dark matter (DM) detection through its annihilation into VHE 
$\gamma$-rays has aroused interest to observe the Galactic Center (GC). H.E.S.S. and MAGIC 
observed the GC, measuring a steady flux consistent with a differential power-law slope of 
$\sim$2.2, up to energies of $\sim$20~TeV with no cutoff (Aharonian et al. 2004; Albert et 
al. 2006c). Within the error circle of the measurement of the central source H.E.S.S.\,J1745-290 
are three compelling candidates for the origin of the VHE emission: the shell-type SNR Sgr~A~East, 
the newly discovered PWN G~359.95$-$0.04, and the supermassive BH Sgr~A$^{\star}$ itself. 
Plausible radiation mechanisms include IC scattering of energetic electrons, the decay of 
pions produced in the interactions of energetic hadrons with the ISM or dense radiation 
fields, and curvature radiation of UHE protons close to Sgr A$^{\star}$. These considerations 
disfavor DM annihilation as the main origin of the detected flux, whereas a more conventional 
astrophysical mechanism is likely to be at work (e.g., Aharonian et al. 2006f). Furthermore, 
the lack of flux variability on hour/day/year timescales suggests that particle acceleration 
occurs in a steady object, such as a SNR or a PWN, and not in the central BH. 

The GC diffuse emission correlates with molecular clouds and suggests an enhanced CR spectrum 
in the Galactic center (Aharonian et al. 2006g). Its morphology and spectrum suggest recent in 
situ CR acceleration: because the photon indexes of the diffuse emission and of the central 
source H.E.S.S.\,J1745-290 are similar, the latter source could be the accelerator in question.

\section{Star-forming galaxies}

Diffuse $\gamma$-ray emission from pp interactions of CR nuclei with target ISM and photons makes up 
$\sim$90\% of the $>$100~MeV luminosity of the Milky Way (Strong et al. 2000). However, the VHE flux 
from a galaxy like the Milky Way located 1 Mpc away would be well below current IACT sensitivities. 
Indeed, only loose upper limits on the VHE flux from normal galaxies have been obtained, even for local 
galaxies and for the VHE-bright starburst galaxies (e.g., Torres et al. 2004). 
%(Only the LMC has been detected, albeit in the softer EGRET range: 
%Sreekumar et al. 1992). 
Detailed models of VHE emission from NGC~253 (V\"olk et al. 1996; Paglione et al. 1996; Domingo-Santamar\'{\i}a 
\& Torres 2005) and for Arp~220 (Torres 2004) are only loosely constrained by current upper limits 
(Aharonian et al. 2005c; Albert et al. 2007e). 
%Detection of the VHE emission associated with ongoing star formation in the universe clearly is one 
%major lingering goal of VHE astrophysics.

\section{Active galactic nuclei}

Supermassive black holes (SMBHs) reside in the cores of most galaxies. The fueling of SMBHs 
by infalling matter produces the spectacular activity observed in active galactic nuclei (AGNs).

The current AGN paradigm includes a central engine, most likely a SMBH, surrounded by an
accretion disk and by fast-moving clouds, which emit Doppler-broadened lines (Urry \& Padovani 
1995; Padovani 2007). In $\sim$10$\%$ of all AGNs, the infalling matter turns on powerful 
collimated jets that shoot out from the SMBH in opposite directions, likely perpendicular 
to the disk, at relativistic speeds. 

If a relativistic jet is viewed at small angle to its axis the observed jet emission is 
amplified by relativistic beaming 
	\footnote{ Defining the relativistic Doppler factor as 
	$\delta$$\equiv$$[\Gamma$$(1$$-$$\beta \cos \theta)]^{-1}$ (with 
	$\beta$$=$$v/c$ the jet speed normalized to the speed of light, 
	$\Gamma$$=$$1/$$\sqrt{(1-\beta^2)}$, and $\theta$ the angle w.r.t. the
	line of sight), the observed and intrinsic luminosities at a given 
	frequency $f$ are related by $L_f^{\rm obs}$$=$$\delta^p$$L_f^{\rm em}$ 
	with $p$$\sim$2-3, and the variability timescales are related by 
	$\Delta t_{\rm obs}$$=$$\delta^{-1}$$\Delta t_{\rm em}$. For 
	$\theta$$\sim$0$^{\circ}$ and $\delta$$\sim$2$\,\Gamma$ the observed 
	luminosity can be amplified by factors $\sim$400--10$^4$ (for, 
	typically, $\Gamma$$\sim$10 and $p$$\sim$2-3); whereas 
	$\theta$$\sim$$1/\Gamma$ implies $\delta$$\sim$$\Gamma$, 
	with a luminosity amplification of $\sim$10$^2$--10$^3$.}
and dominates the observed emission. Such sources are called blazars. Given the blazars' compactness 
(as suggested by their short variability timescales), all GeV/TeV photons would be absorbed through 
pair-producing $\gamma\gamma$ collisions with target X-ray/IR photons. Beaming ensures the intrinsic 
radiation density to be much smaller than the observed one, so that $\gamma$-ray photons encounter a 
much lower $\gamma\gamma$ opacity and hence manage to leave the source: reversing the argument, 
$\gamma$-ray detection is a proof of strongly anisotropic (e.g., beamed) emission.

The spectral energy distributions (SEDs) of blazars are generally characterized by two broad 
humps, peaking at, respectively, IR/X-ray and GeV-TeV frequencies (Ulrich et al. 1997). Analyses of 
blazar SEDs (Fossati et al. 1998; Ghisellini et al. 1998) have suggested that: 
{\it (i)} higher/lower-luminosity objects have both humps peaking at lower/higher frequencies 
(they are called, respectively, LBLs and HBLs); 
{\it (ii)} the luminosity ratio between the high- and low-frequency bumps increases with luminosity; 
{\it (iii)} at the highest luminosities the $\gamma$-ray output dominates the total luminosity. 
%In this proposed scheme, BL Lac objects populate the lower luminosity part of the 
%sequence (\cite{g+98}).

The mainstream interpretation of the blazars SEDs is synchrotron-Compton emission, i.e. synchrotron 
emission (peaked in the IR/X-ray range) from a time-varying population of ultra-relativistic electrons 
moving in a strong magnetic field, and IC emission (peaked in the $\sim$100$\,$MeV--100$\,$GeV range) 
from soft photons scattering off energetic electrons. Depending on the relative efficiency of the 
relativistic particles' cooling through scattering with photon fields that are internal to jet or external 
to it, the synchrotron and Compton components peak at, respectively, UV/X-ray and GeV--TeV energies 
(synchrotron-self-Compton [SSC] scheme: e.g., Maraschi et al. 1992) or at IR/optical and MeV--GeV energies 
(external-IC [EIC] scheme, see Dermer \& Schlickeiser 1993). Hybrid SSC/EIC models have also been proposed 
(Ghisellini 1999). Alternative models of VHE emission involve, e.g., two electron populations, one 
--primary-- accelerated within the jet and the other --secondary-- generated by electromagnetic cascades 
initiated by primary protons/nuclei that had been accelerated in the jet (Mannheim 1993); or a population 
of extremely energetic protons emitting by synchrotron radiation (Aharonian 2000). 

The emitting particles are accelerated within the relativistic jets which transport energy from the 
central SMBH outwards (Rees 1967). In the popular SSC framework this process is approximated with a 
series of relativistically moving homogeneous regions (blobs), where particle acceleration and 
radiation take place (e.g., Maraschi et al. 1992). The X-ray and $\gamma$-ray emission, with its 
extremely fast and correlated multi-frequency variability, indicates that often a single region dominates 
the emission.

VHE data are of crucial importance to close the SSC model. Even in the simplest one-zone SSC model of 
blazar emission, knowledge of the whole SED up to the VHE regime is required for a complete description 
of the emitting electrons' distribution and environment (e.g., Tavecchio et al. 1998). The parameters 
that specify the properties of the emitting plasma in the basic SSC model are: the electron distributions 
normalization, low- and high-energy slopes, and min/break/max energy, and the plasma blobs magnetic field, 
size and Lorentz factor. Knowing only the IR/X-ray peak would give info on the shape (i.e., the 
broken-power-law slopes) of the electron distribution but would leave all other parameters unconstrained 
(e.g., Tavecchio et al. 1998). However, accurate knowledge of blazar emission mechanism(s) requires 
{\it simultaneous} broadband $\gamma$-ray and X-ray (i.e., IC and synchrotron) data. In fact, a 
simultaneous SED can act as a snapshot of the emitting population of particles at a given time. 

Blazar observations have been a top priority for VHE astrophysics ever since the discovery of TeV 
emission from Mrk~421 (Punch et al. 1992). To date, firm blazar TeV detections include 15 HBLs and 1 LBL. 
(One further non-blazar AGN, M~87, has also been detected, see Aharonian et al. 2006h).) 
%Such detections are much fewer than GeV ones ($\magcir$130, see Padovani 2007), 
%the effect being possibly blamed on cosmological absorption of TeV photons (see below). 

The known TeV blazars are variable in flux in all wavebands. Even simple one-zone homogeneous SSC 
modeling predicts the X-ray and TeV flux variability to be closely correlated, both emissions being 
linked to the same electron population. Observational evidence, although still statistically limited, 
supports this prediction (e.g., Pian et al. 1998). 
(Some TeV flares that show no simultaneous X-ray counterpart may be explained as IC radiation 
from an additional, very compact and dense, electron population -- see Krawczynski et al. 2004.)
Blazar variability, in flux and spectrum, has been observed at VHE frequencies down to minute timescales. 
For Mkn~501, observed with the MAGIC telescope at $>$100~GeV during 24 nights between May and July 2005, 
the integrated flux and the differential photon spectra could be measured on a night-by-night basis 
(Albert et al. 2007f). 
During the observational campaign, the flux variations (from $\sim$0.4 to $\sim$4 crab units) were 
correlated with the spectral changes (i.e., harder spectra for higher fluxes), and a spectral peak 
showed up during the most active phases. A rapid flare occurred on the night of 10 July 2005, showing 
a doubling time as short as $\sim$2 minutes and a delay of $\sim$3 minutes as a function of energy of the 
emitted photons. 

One further aspect of TeV spectra of blazars is that they can be used as probes of the 
Extragalactic Background Light (EBL), i.e. the integrated light arising from the evolving 
stellar populations of galaxies (see Hauser \& Dwek 2001). The TeV photons emitted by a blazar 
interact with the EBL photons and are likely absorbed via pair production. Whatever its intrinsic 
shape at emission, after travelling through the EBL-filled space, a blazar spectrum will reach 
the observer distorted by absorption. This effect, which is stronger for more distant objects (e.g., 
Stecker et al. 1992), is the most likely origin of the avoidance zone (i.e., no flat spectra at 
high redshift) in the observed spectral slope vs redshift plot. The strength of the absorption is 
measured by $\tau_{\gamma \gamma}(E)$, the optical depth for attenuation between the blazar, 
located at a distance $D(z)$, and the Earth (Fazio \& Stecker 1970; Stecker et al. 1992). 
Usually, either {\it (i)} the shape and intensity of EBL($z$) is assumed, and the TeV spectrum 
is corrected before the SSC modeling is performed (e.g., de Jager \& Stecker 2002; Kneiske et al. 2004); 
or {\it (ii)} based on assumptions on the intrinsic VHE spectrum, EBL($z$) is solved for: 
e.g., based on analysis of the observed hard VHE spectra of the distant blazars 1ES~1101-232 and 
H~2359-309, a low EBL energy density at $z$$\mincir$0.2 has been derived (Aharonian et al. 2006i). The two 
approaches can be used in combination to estimate the distance to the VHE source (Mazin \& Goebel 2007).

\section{Gamma-ray bursts}

There is a prevailing consensus that the basic mechanism of GRB emission is an expanding relativistic 
fireball (Rees \& Meszaros 1992, Meszaros \& Rees 1993, Sari et al. 1998), with the beamed radiation 
($\delta$$\sim$O(10$^2$) due to internal/external shocks (prompt/afterglow phase, respectively). If so, 
the emitting particles (electrons and/or protons) are accelerated to very high energies. 

In the fireball shock framework, several models have predicted VHE emission during both the prompt and 
afterglow phases of the GRB (e.g., Meszaros 2006). This can occur as a result of electron 
self-IC emission from the internal shock or the external forward/reverse shock. Seed photons can be produced 
locally (through synchrotron, or be the leftover of the initial radiation content responsible for the 
acceleration of the fireball) or can be produced in, e.g., the shell of a previously exploded SN. In the 
latter case, the SN photons may also act as targets for the $\gamma \gamma$ absorption, and in this case 
the VHE emission could be severely dimmed. If the emission processes are indeed synchrotron and IC, then 
a blazar-like SED is predicted, with a double-peak shape extending into the VHE band. In such theoretical 
freedom, VHE observations of GRBs could help constraining GRB models. 

MAGIC observed part of the prompt-emission phase of GRB~050713a as a response to an alert by the Swift 
satellite (Albert et al. 2006d)
%
%\footnote{Being a member of the GRB Coordinate Network (GCN), and exploiting 
%	    its relatively low threshold energy and its short slewing time 
%	   ($\mincir$20~s), MAGIC has the unique capability to point at the GRB 
%	    location still during the prompt-emission phase and measure TeV 
%	    emission.}.
%
However, no excess at $>$175~GeV was detected, neither during the prompt emission phase nor later 
-- but the upper limits to the MAGIC flux are compatible with simple extrapolations of the Swift 
$\Gamma$$\simeq$1.6 power-law spectrum to hundreds of GeV. In general, however, the cosmological 
distances of these sources prevent VHE detection (see Albert et al. 2007g): the average redshift 
of the GRBs for which MAGIC was alerted (and whose $z$ are known) is $\bar z$$=$3.22, whereas at 
70~GeV the cosmological $\gamma$-ray horizon is $z$$\sim$1. Complementary Whipple data (Horan et 
al. 2007) and MILAGRO data (Abdo et al. 2007) provide upper limits on, respectively, the late VHE 
emission ($\sim$4 hr after the burst) from several long-duration GRBs, and on the prompt/delayed 
emission from several, reputedly nearby ($z$$\mincir$0.5), short-duration GRBs.

%\section{Exotica: Quantum gravity}
%
%Blazar emission allows to explore non-conventional physics. Different approaches to 
%quantum gravity (QG) lead to similar quantifications of the first-order effects of a 
%violation induced in the Lorentz-Poincar\'e symmetry. Such violation should cause a 
%dependence of the speed of light on the photon energy, $E$, as $c^{\prime}=c 
%(1-\xi{E \over E_{\rm QG}}$, where $E_{\rm QG}$$\sim$$10^{19}$ GeV (i.e., 
%the Planck scale), and $\xi$$=$$\pm 1$ depending on the dynamical framework (e.g., 
%Amelino-Camelia et al. 1998). Consequently, two photons of energy $E$ 
%and $E+\Delta E$, emitted simultaneously at a distance $L$, would be separated at 
%arrival by a time delay $\Delta t = \xi {\Delta E \over E_{\rm QG}}{L \over c}$. 
%Short arrival delay times (measured within minute-long timescale variability), large 
%source distances, and large photon energy differences can lead to significant 
%astrophysical (lower) limits of the symmetry-breaking scale. An energy-dependent delay 
%of the peak emission during minute-timescale flux variability in Mrk~501 implies, if 
%interpreted in this framework, $E_{\rm QG}$$\magcir$0.6$\times$10$^{17}$~GeV \cite{mk501}
%
%Amelino-Camelia, G. et al. 1998, Nature, 393, 763

\section{Dark matter}

Evidence for departure of cosmological motions from the predictions of Newtonian 
dynamics based on visible matter, interpreted as due to the the presence of DM, are 
well established -- from galaxy scales (e.g., van Albada et al. 1985) to 
galaxy-cluster scales (e.g., Sarazin 1986) to cosmological scales (e.g., Spergel et al. 2003). 

DM particle candidates should be weakly interacting with ordinary matter (and hence neutral). 
The theoretically favored ones, which are heavier than the proton, are dubbed weakly interacting 
massive particles (WIMPs). WIMPs should be long-lived enough to have survived from their decoupling 
from radiation in the early universe into the present epoch. Except for the neutrino, which is the 
only DM particle known to exist within the Standard Model of elementary particles (with a relic 
background number density of $\sim$50~ cm$^{-3}$ for each active neutrino species) but which is too 
light ($m_\nu$$\mincir$1~eV) to contribute significantly to $\Omega_m$ given the current cosmological 
model, WIMP candidates have been proposed only within theoretical frameworks mainly motivated by 
extensions of the Standard Model of particle physics (e.g., the R-parity conserving supersymmetry 
[SUSY]). Among current WIMP candidates (see Bertone et al. 2005), the neutralino, which is the 
lightest SUSY particle, is the most popular candidate. Its relic density is compatible with {\it W}MAP 
bounds (see Munoz 2004).

WIMPs could be detected directly, via elastic scattering vith targets on Earth, or 
indirectly, by their self-annihilation products in high-density DM environments. 
DM annihilation can generate $\gamma$-rays through several processes. Most distinctive are 
those that result in mono-energetic spectral lines, $\chi\chi$$\rightarrow$$\gamma\gamma$, 
$\chi\chi$$\rightarrow$$\gamma$$Z$ or $\chi\chi$$\rightarrow$$\gamma$$h$. However, in most 
models the processes only take place through loop diagrams; hence the cross sections for 
such final states are quite suppressed, and the lines are weak and experimentally 
challenging to observe. A continuum $\gamma$-ray spectrum can also be produced through 
the fragmentation and cascades of most other annihilation products. The resulting spectral 
shape depends on the dominant annihilation modes (see Bergstr\"om \& Hooper 2006), whereas 
the normalization depends on the WIMP's mass and velocity-averaged annihilation cross section 
as well as on the DM density profile. 

Once the astroparticle model has been chosen (e.g., Bergstr\"om et al. 1998), the biggest 
uncertainties are of astrophysical nature. Superposed to any VHE emission from the 
decaying DM (cosmological, non-baryonic signal), galaxies can display a VHE emission from astrophysical 
sources associated with the visible matter distribution (astrophysical, baryonic signal). 
The ratio of the former to the latter is maximized in small, low-$L$, low-SFR galaxy. 
This is because the dark-to-visible mass ratio as well as the central DM density increase 
with decreasing luminosity (Persic et al. 1996). Clearly, distance dilution of the signal opposes detection, 
so galaxies candidate for indirect DM detection should be chosen among nearby objects. In 
conclusion, the best obserational targets for DM detection are the Milky Way dwarf spheroidal 
galaxies (e.g., Draco, Sculptor, Fornax, Carina, Sextans, Ursa Minor). A further issue, 
stemming from the $\rho^2$ dependence (as a result of annihilation) of the normalization integral 
of the $\gamma$-ray emission, concerns the shape of the inner halo profile, i.e. whether the latter 
is cuspy or cored. Cuspy profiles are produced in cosmological N-body simulations of halo formation 
(Navarro et al. 1997), whereas cored profiles are suggested by the measured rotation curves of disk 
galaxies (Borriello \& Salucci 2001) -- also in low-surface-brightness galaxies, 
where the local self-gravity of baryons is virtually negligible (de Blok et al. 2001). 

These considerations (and uncertainties) have been incorporated in detailed predictions of the 
$\gamma$-ray flux expected from the annihilation of the neutralino pairs. Outlooks for VHE neutralino 
detection in Draco by current IACTs are not very promising: for a neutralino mass $m_\chi$$=$100~GeV 
and a variety of annihilation modes, and in the favorable case of a maximal (cuspy) inner halo profile, 
VHE detection (by MAGIC in 40hr observation) can occur if average value of the neutralino's cross 
section times velocity is $<$$\sigma$$v$$>$$\magcir$10$^{-25}$~cm$^3$s$^{-1}$, which is somewhat 
larger than the maximum value for a thermal relic with a density equal to the measured (cold) DM 
density (but may be fine for non-thermally generated relics) in the allowed SUSY parameter space. 
The prospects are better in the HE range (100~MeV--10~GeV): for a maximal (cuspy) halo, 1 yr 
of GLAST observation should be able to yield a detection if $m_\chi$$\mincir$500~GeV and 
$<$$\sigma$$v$$>$$\sim$3$\times$10$^{-25}$~cm$^3$s$^{-1}$ (Bergstr\"om \& Hooper 2006).
%S\'anchez-Conde et al. 2007).

No evidence of DM annihilation $\gamma$-rays has been unambiguously claimed so far. An 
apparently extended signal from the direction of NGC~253 
%, originally claimed by the CANGAROO collaboration 
%(Itoh et al. 2002) and attributed to the halo of NGC~253 (Itoh et al. 2003a) as arising from a 
%combination of astrophysical emission (Itoh et al. 2003b) and DM-annihilation emission 
%(Itoh et al. 2003c), was later 
has been definitely interpreted as due to hardware malfunction (Itoh et al. 2007).

\begin{acknowledgements}
We acknowledge the MAGIC collaboration for providing a stimulating, friendly, and effective working 
environment. MP thanks Pasquale Blasi for inviting this review, which was presented at the 51th general 
meeting of the Italian Astronomical Society (Florence, Apr. 17-20, 2007).
\end{acknowledgements}
\bigskip

\def\ref{\par\noindent\hangindent 20pt}

\noindent
{\bf References}
\vglue 0.2truecm

\ref{\small Abdo, A.A., et al. 2007, arXiv:07051554}
\ref{\small Aharonian, F.A. 2000, New Astron., 5 , 377}
\ref{\small Aharonian, F.A., et al. 2007, A\&A, 464, 235}
\ref{\small Aharonian, F.A., et al. 2006a, ApJ, 636, 777}
\ref{\small Aharonian, F.A., et al. 2006b, A\&A, 449, 223}
\ref{\small Aharonian, F.A., et al. 2006c, A\&A, 448, L43}
\ref{\small Aharonian, F.A., et al. 2006d, A\&A, 460, 365}
\ref{\small Aharonian, F.A., et al. 2006e, A\&A, 460, 743}
\ref{\small Aharonian, F.A., et al. 2006f, Phys. Rev. Lett., 97, 221102}
\ref{\small Aharonian, F.A., et al. 2006g, Nature, 439, 695}
\ref{\small Aharonian, F.A., et al. 2006h, Science, 314, 1424}
\ref{\small Aharonian, F.A., et al. 2006i, Nature, 440, 1018}
\ref{\small Aharonian, F.A., et al. 2005a, A\&A, 442, 1}
\ref{\small Aharonian, F.A., et al. 2005b, Science, 309, 746}
\ref{\small Aharonian, F.A., et al. 2005c, A\&A, 442, 177}
\ref{\small Aharonian, F.A., et al. 2004, A\&A, 425, L13}
\ref{\small Albert, J., et al. 2007a, arXiv:0705.3119}
\ref{\small Albert, J., et al. 2007b, astro-ph/0702077}
\ref{\small Albert, J., et al. 2007c, arXiv:0705.3244}
\ref{\small Albert, J., et al. 2007d, arXiv:0706.1505}
\ref{\small Albert, J., et al. 2007e, ApJ, 658, 245}
\ref{\small Albert, J., et al. 2007f, astro-ph/0702008}
\ref{\small Albert, J., et al. 2007g, astro-ph/0612548}
\ref{\small Albert, J., et al. 2006a, ApJ, 643, L53}
\ref{\small Albert, J., et al. 2006b, Science, 312, 1771}
\ref{\small Albert, J., et al. 2006c, ApJ, 638, L101}
\ref{\small Albert, J., et al. 2006d, ApJ, 641, L9}
\ref{\small Berezhko, E.G., \& V\"olk, H.J. 2006, A\&A, 451, 981}
\ref{\small Berezhko, E.G., et al. 2003, A\&A, 400, 971}
\ref{\small Bergstr\"om, L., \& Hooper, D. 2006, Phys. Rev. D, 73, 063510}
\ref{\small Bergstr\"om, L., et al. 1998, Astropart. Phys., 9, 137}
\ref{\small Bertone, G., et al. 2005 Phys. Rep. 405, 279}
%\ref{\small Blandford, R., \& Eichler, D. 1987, Phys.Rep., 154, 1}
\ref{\small Blasi, P. 2005, Mod. Phys. Lett. A20, 3055}
\ref{\small Borriello, A. \& Salucci, P. 2001, MNRAS, 323, 285}
\ref{\small Cheng, K.S., et al. 1986, ApJ, 300, 500}
\ref{\small Daugherty, J.K., \& Harding, A. 1982, ApJ, 252, 337}
\ref{\small Daugherty, J.K., \& Harding, A. 1996, ApJ, 458, 278}
\ref{\small de~Bloek, W.J.G., et al. 2001, ApJ, 552, L23}
\ref{\small de~Jager, O.C., \& Stecker, F.W. 2002, ApJ, 566, 738}
\ref{\small Dermer, C., \& Schlickeiser, R. 1993, ApJ, 416, 458}
\ref{\small Domingo-S., E., \& Torres, D. 2005, A\&A, 444, 403}
\ref{\small Fazio, G.G., \& Stecker, F.W. 1970, Nature, 226, 135}
\ref{\small Fossati, G., et al. 1998, MNRAS, 299, 433}
\ref{\small Gallant, Y. 2007, astro-ph/0611720}
\ref{\small Ghisellini, G., et al. 1998, MNRAS, 301, 451}
\ref{\small Harding, A.K., et al. 2005, ApJ, 622, 531}
\ref{\small Hauser, M.G., \& Dwek, E. 2001, ARA\&A, 39, 249}
\ref{\small Horan, D., et al. 2007, astro-ph/0701281}
%\ref{\small Itoh, C., et al. 2002, A\&A, 396, L1}
%\ref{\small Itoh, C., et al. 2003a, A\&A, 402, 443}
%\ref{\small Itoh, C., et al. 2003b, ApJ, 584, L65}
%\ref{\small Itoh, C., et al. 2003c, ApJ, 596, 216}
\ref{\small Itoh, C., et al. 2007, A\&A, 462, 67}
\ref{\small Kneiske, T.M., et al. 2004, A\&A, 413, 807}
\ref{\small Krawczynski, H., et al. 2004, ApJ, 601, 151}
\ref{\small Mannheim, K. 1993, A\&A, 269, 76}
\ref{\small Maraschi, L, et al. 1992, ApJ, 397, L5}
\ref{\small Mazin, D., \& Goebel, F. 2007, ApJ, 655, L13}
\ref{\small Meszaros, P. 2006, Rep. Prog. Phys., 69, 2259}
\ref{\small Meszaros, P., \& Rees, M.J. 1993, ApJ, 405, 278}
\ref{\small Munoz, C. 2004, Int.J.Mod.Phys. A, 19, 3093}
\ref{\small Navarro, J.F., et al. 1997 ApJ, 490, 493}
\ref{\small O\~na-Wilhelmi, E., et al. 2007, ICRC07 abstract 528}
\ref{\small Padovani, P. 2007, arXiv:0704.0184 }
%\ref{\small Padovani, P. 2007, in First GLAST Symposium, Stanford U. (arXiv:0704.0184) }
\ref{\small Paglione, T.A.D., et al. 1996, ApJ, 460, 295}
\ref{\small Persic, M., et al. 1996, MNRAS, 281, 27}
\ref{\small Pian, E., et al. 1998, ApJ, 492, L17}
\ref{\small Porter T.A. et al. 2006, ApJ, 648, L29}
\ref{\small Punch, M., et al. 1992, Nature, 358, 477}
\ref{\small Rees, M.J. 1967, MNRAS, 137, 429}
\ref{\small Rees, M.J., \& Meszaros, P. 1992, MNRAS, 258, 41p}
\ref{\small Romani, R.W. 1996, ApJ, 470, 469}
%\ref{\small S\'anchez-Conde, M.A., et al. 2007, JCAP submitted (astro-ph/0701426)}
\ref{\small Sarazin, C. 1986, Rev. Mod. Phys., 58, 1}
\ref{\small Sari, R., et al. 1998, ApJ, 497, L17}
\ref{\small Spergel, D.N., et al. 2003, ApJS, 148, 175}
%\ref{\small Sreekumar, P., et al. 1992, ApJ, 400, L67 LMC}
\ref{\small Stecker, F.W., et al. 1992, ApJ, 390, L49}
\ref{\small Strong, A.W., et al. 2000, ApJ, 537, 763}
\ref{\small Tavecchio, F., et al. 1998, ApJ, 509, 608}
\ref{\small Torres, D.F., et al. 2003, Phys. Rep., 382, 303}
\ref{\small Torres, D.F. 2004, ApJ, 617, 966}
\ref{\small Torres, D.F., et al. 2004, ApJ, 2004, 607, L99}
\ref{\small Ulrich, M.-H., et al. 1997, ARA\&A, 35, 445}
\ref{\small Urry, C.M., \& Padovani, P. 1995, PASP, 107, 803}
\ref{\small van Albada, T.S., et al. 1985, ApJ, 295, 305}
\ref{\small V\"olk, H., et al. 1996, Space Sci. Rev., 75, 279 }
\ref{\small Wang, X.Y., et al. 2006, ApJ, 641, L89}
\ref{\small Weekes, T.C., et al. 1989, ApJ, 342, 379}

\end{document}